\documentclass[prl,a4paper]{revtex4}

\usepackage{amsmath,amssymb}
\usepackage{epsfig}

\newcommand{\numero}[1]{\noindent #1.~}

\begin{document}
\title{Geometric scaling as traveling waves}

\author{S. Munier%
\footnote{Permanent address after October, 1: 
Centre de physique th{\'e}orique,
{\'E}cole polytechnique, 91128 Palaiseau cedex, France.}
}\email{Stephane.Munier@cpht.polytechnique.fr} 
\author{R. Peschanski}
\email{pesch@spht.saclay.cea.fr}
\affiliation{Service de physique th{\'e}orique, CEA/Saclay,
  91191 Gif-sur-Yvette cedex, France\footnote{%
URA 2306, unit\'e de recherche associ\'ee au CNRS.}}

\begin{abstract}
We show the relevance of the nonlinear
Fisher and Kolmogorov-Petrovsky-Piscounov (KPP) equation
to the problem of high energy evolution of the QCD amplitudes.
We explain how the traveling wave solutions of this equation
are related to geometric scaling,
a phenomenon observed in deep-inelastic scattering experiments.
Geometric scaling is for the first time shown to result from an
exact solution of nonlinear QCD evolution equations.
Using general results on the KPP equation,
we compute the velocity of the wave front, which gives
the full high energy dependence of the saturation scale.
\end{abstract}

\maketitle

\numero{1}Geometric scaling \cite{Stasto:2000er} is an 
interesting phenomenological feature
of high energy deep-inelastic scattering (DIS).
It is expressed as a scaling property of the virtual photon-proton 
cross section, namely
\begin{equation}
\sigma^{\gamma^*}(Y,Q)=\sigma^{\gamma^*}\!
\left(\frac{Q^2}{Q_s^2(Y)}\right)\ ,
\end{equation}
where $Q$ is the virtuality of the photon,
$Y$ the total rapidity and $Q_s$ an increasing function of $Y$ called the
saturation scale \cite{endnote17}.

It is convenient to work within the QCD dipole picture of DIS
\cite{Nikolaev:1991ja}.
In the leading logs ($Y$) approximation of perturbative QCD, 
the cross section factorizes as
\begin{equation}
\sigma^{\gamma^*}(Y,Q)=\int_0^\infty {x}_{01}\,dx_{01}\int_0^1 dz\,
|\psi(z,{x}_{01}Q)|^2\, N(Y,{x}_{01})\ .
\label{eq:gammaN}
\end{equation}
$\psi(z,{x}_{01}Q)$ is the photon wave function 
on a $q\bar q$ dipole of 
size $x_{01}$, and $z$ is the longitudinal momentum fraction of the photon
carried by the quark.
$N(Y,{x}_{01})$ is the dipole forward 
scattering amplitude.
Let us define
\begin{equation}
{\cal N}(Y,k)=\int_0^{\infty} \frac{dx_{01}}{x_{01}}
J_0(kx_{01})\,N(Y,x_{01})\ .
\label{eq:fourier}
\end{equation}
In this picture, the geometric scaling property reads
\begin{equation}
{\cal N}(Y,k)={\cal N}\left(\frac{k^2}{Q_s^2(Y)}\right)\ .
\label{eq:defscaling}
\end{equation}
Within suitable approximations
(large $N_c$, summation of fan diagrams, spatial homogeneity) 
and starting from the Balitsky-Kovchegov equation \cite{Balitsky},
it has been shown \cite{Kovchegov} that this quantity obeys
the nonlinear evolution equation
\begin{equation}
{\partial_Y}{\cal N}=\bar\alpha
\chi\left(-\partial_L\right){\cal N}
-\bar\alpha\, {\cal N}^2\ ,
\label{eq:kov}
\end{equation}
where $\bar\alpha=\alpha_s N_c/\pi$,
$\chi(\gamma)=2\psi(1)-\psi(\gamma)-\psi(1\!-\!\gamma)$ is the
characteristic function of the BFKL kernel \cite{Lipatov:1976zz},
$L=\log (k^2/k_0^2)$ and $k_0$ is some fixed low momentum scale.
It is well-known that the BFKL kernel can be expanded 
to second order around $\gamma\!=\!{\scriptstyle \frac 12}$,
if one sticks to the kinematical regime $8\bar\alpha Y \gg L$. 
We expect this commonly used
approximation to remain valid 
for the full nonlinear equation~(\ref{eq:kov}).
The latter boils down to a parabolic nonlinear
partial derivative equation:
\begin{equation}
{\partial_Y}{\cal N}=\bar\alpha
\bar\chi\left(-\partial_L\right){\cal N}
-\bar\alpha\, {\cal N}^2\ ,
\label{eq:kovchegov}
\end{equation}
with
\begin{equation}
\bar\chi\left(-\partial_L\right)
=\chi\left({\scriptstyle \frac12}\right)+
\frac{\chi^{\prime\prime}\!\left({\scriptstyle\frac12}\right)}{2}
\left(\partial_L+{\frac12}\right)^2\ .
\label{eq:expansion}
\end{equation}
The question we want to address
is whether there are exact asymptotic (in $Y$)
solutions of Eq.(\ref{eq:kovchegov}) exhibiting geometric scaling.\\

%%%%%%%%%%%%%%%%%%%%%%%%%%%%%%%%%%%%%%%%%%%%%%%%%%%%%%%%%%

\numero{2}Introducing the notation 
$\omega=\chi({\scriptsize{\frac12}})$, 
$D=\chi^{\prime\prime}({\scriptsize{\frac12}})$, 
and defining $\bar\gamma=1-{\scriptstyle\frac12}\sqrt{1+8\omega/D}$,
the change of variables
\begin{equation}
\begin{split}
t&=\frac{\bar\alpha D}{2}(1\!-\!\bar\gamma)^2\,Y\ \ \ ,\ \ \  
x=(1\!-\!\bar\gamma)
\left(L+\frac{\bar\alpha D}{2}\,Y\right)\ ,\\
u(t,x)&=\frac{2}{D(1\!-\!\bar\gamma)^2}
\,{\cal N}\left(\frac{2t}{\bar\alpha D(1\!-\!\bar\gamma)^2}\,,\,
\frac{x}{1\!-\!\bar\gamma}
-\frac{t}{(1\!-\!\bar\gamma)^2}
\right)
\end{split}
\label{eq:map}
\end{equation}
brings Eq.(\ref{eq:kovchegov}) for $\cal N$ into 
the Fisher and Kolmogorov-Petrovsky-Piscounov (KPP) equation
\cite{KPP} for $u$:
\begin{equation}
\partial_t u(t,x)=\partial_x^2 u(t,x)+u(t,x)(1-u(t,x))\ .
\label{eq:KPP}
\end{equation}

%%%%%%%%%%%%%%%%%%%%%%%%%%%%%%%%%%%%%%%%%%%%%%%%

\begin{figure}
\epsfig{file=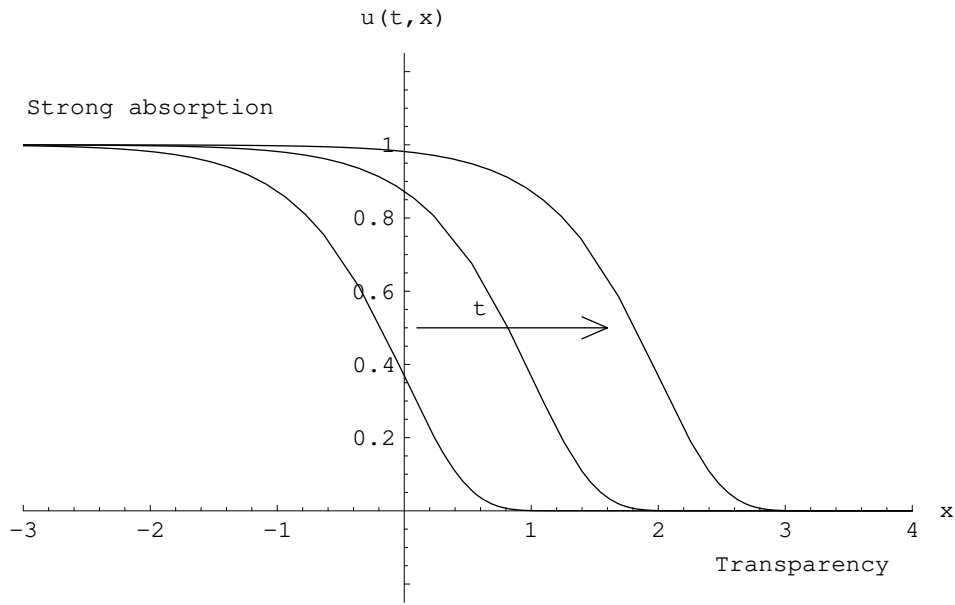,width=14cm}
\caption{\label{fig:fkpp}{\it Typical traveling wave solution.}
The function $u(t,x)$ is
represented for three different times. The wave front connecting the regions
$u=1$ and $u=0$ travels
from the left
to the right as $t$ increases. That illustrates 
how the ``strong absorption''
region invades the ``transparency'' region (see text).}
\end{figure}

We are going to use some general results on the existence
of traveling wave solutions to the KPP equation \cite{Bramson}.
If one chooses an initial condition at time $t=t_0$ such that 
$u(t_0,x)$ decreases smoothly from 1
to 0 as $x$ goes from $-\infty$ to $+\infty$, and 
has the asymptotic behavior
\begin{equation}
u(t_0,x)\underset{x\rightarrow+\infty}{\sim} e^{-\beta x}\ ,
\label{eq:initialcond}
\end{equation}
it is proven \cite{Bramson} that the KPP equation admits
traveling wave solutions at large times. This means that 
there exists a function 
of one variable $w$ such that 
\begin{equation}
u(t,x)\underset{t\rightarrow +\infty}{\sim}
w(x-m_\beta(t))
\end{equation}
uniformly in $x$.
Such a solution is depicted on Fig.(\ref{fig:fkpp}).
The function $m_\beta(t)$ depends on the initial condition:
\begin{equation}
\begin{aligned}
m_\beta(t)&=c(\beta)t+{\cal O}(1) 
	& \mbox{for $\beta<\beta_c$}\ ,\\
m_\beta(t)&=2 t-{\scriptstyle \frac12}\log t+{\cal O}(1)
	& \mbox{for $\beta=\beta_c$}\ ,\\
m_\beta(t)&=2 t-{\scriptstyle \frac32}\log t+{\cal O}(1)\ \ \ 
	& \mbox{for $\beta>\beta_c$}\ ,
\end{aligned}
\label{eq:velocity}
\end{equation}
where $c(\beta)=\beta+1/\beta$, and 
the critical value
$\beta_c=1$ corresponds to the minimum of $c(\beta)$.
Note that for critical or supercritical $\beta\ge\beta_c$, the
velocity of the 
traveling wave front $dm_\beta(t)/dt$ is independent of $\beta$ 
up to a constant.\\

%%%%%%%%%%%%%%%%%%%%%%%%%%%%%%%%%%%%%%%%%%%%%%%%%

\numero{3}Let us show that the initial conditions 
for QCD are of the form of Eq.(\ref{eq:initialcond}).
For an initial value $Y=Y_0$, corresponding to 
$t=t_0=\bar\alpha D(1-\bar\gamma)^2 Y_0/2$ and
for large $L$, corresponding to large $x$,
the behaviour of ${\cal N}$ 
is given by perturbative
QCD. Asymptotically in $L$ ({\it i.e.} in $x$ at
fixed $t$), it reads
\begin{equation}
{\cal N}\propto {k^{-2\gamma_0}}\ ,\ \gamma_0=1\ .
\label{eq:physic}
\end{equation}
We get the relationship between $\beta$ and $\gamma$ from the 
coordinate map~(\ref{eq:map}):
$\beta=\gamma_0/({1-\bar\gamma})$.
In particular, $u(t_0,+\infty)\propto 
{\cal N}(Y_0,L\!\rightarrow\!+\infty)=0$. Formula~(\ref{eq:physic}) 
through the Fourier transform~(\ref{eq:fourier}), corresponds
to QCD color transparency, namely $N(Y,x_{01})\propto x_{01}^2$.

The condition $u(t_0,-\infty)=1$ corresponds to the normalization condition
${\cal N}(Y_0,L\rightarrow-\infty)=D(1\!-\!\bar\gamma)^2/2$ 
when $k/k_0\rightarrow 0$.
This infrared limit is not attainable in
perturbative QCD, but it is reasonable
to expect that some upper bound on ${\cal N}$ exist
because of strong absorption.

Interestingly enough,
$\beta$ happens to be equal to $2/\sqrt{1+8\omega/D}=1.55276...$, 
{\it i.e.} it lies within the supercritical regime $\beta>\beta_c$
of the KPP equation.
Note that in this regime, possible logarithmic
prefactors in Eq.(\ref{eq:physic}) would not modify the traveling
wave solution, at variance with the critical case $\beta=\beta_c$ 
\cite{Bramson}. Hence color transparency is the only relevant property
for defining the initial condition at large $L$.

%%%%%%%%%%%%%%%%%%%%%%%%%%%%%%%%%%%%%%%%%%%%%%%%%%%%%%%%%%%%%

According to the above-mentioned mathematical framework, 
traveling wave solutions to 
Eq.(\ref{eq:kovchegov}) for ${\cal N}$ exist
at large~$Y$.
We shall show that they correspond to specific
geometric scaling solutions.
Indeed, the velocity of the traveling wave front is given
by the third equation in~(\ref{eq:velocity}) because we are in the
supercritical regime. 
This means that the large time
solutions depend on a single variable $x-2t+{\scriptstyle\frac32}\log t$. 
From the coordinate mapping~(\ref{eq:map}) ,
one sees that we may write the solution as
\begin{equation}
{\cal N}(Y,k)={\cal N}\left(\frac{k^2}{Q_s^2(Y)}
\right)\ \ \ \mbox{where}\ \ \ 
Q_s^2(Y)=k_0^2\,{Y^{-\frac{3}{2(1-\bar\gamma)}}}
e^{\bar\alpha D(\frac12-\bar\gamma)Y}\ .
\label{eq:geoscale}
\end{equation}
We have absorbed all ${\cal O}(1)$ constants into $k_0^2$.
Eq.(\ref{eq:geoscale}) is nothing else than
geometric scaling, $Q_s(Y)$ being the saturation scale,
{\it cf.} Eq.(\ref{eq:defscaling}). \\

%%%%%%%%%%%%%%%%%%%%%%%%%%%%%%%%%%%%%%%%%%%%%%%%%%%%%%%

\numero{4}Equation (\ref{eq:geoscale}) is the exact asymptotic
solution of the nonlinear Eq.(\ref{eq:kovchegov}) 
and thus geometric
scaling appears as a universal property of this kind of equations.
We obtain the saturation scale including the complete 
prefactor up to factors of order ${\cal O}(1)$.
It comes from a non-trivial mathematical property
of nonlinear equations.
To our knowledge, it is the first exact result taking
into full account the nonlinearities of a QCD evolution 
equation~(\ref{eq:kovchegov}).

Comparing our results to previous studies,
the exponential growth of the saturation scale 
with the rapidity~$Y$ matches the estimates obtained from
the extrapolation of the solution of the linear BFKL equation
up to the saturation scale
\cite{Iancu:2002tr}. The powerlike prefactor that we get and
prove to be universal, matches the one obtained in
Ref.\cite{Mueller:2002zm}, where absorption
was treated as a boundary condition on the linear equation
\cite{endnote18}.

Several numerical studies of Eq.(\ref{eq:kov}) have been performed.
Evidence for ``soliton-like'' solutions \cite{Braun:2000wr} 
and for geometric scaling has been found
\cite{Braun:2000wr,numerics}. Recently, 
formula~(\ref{eq:geoscale}) was proved
to be very well verified \cite{Albacete:2003iq}.
This matches the mathematical properties we have discussed.

Among pending mathematical questions, it would be interesting to have
a similar approach for the original equation~(\ref{eq:kov}), beyond
the approximation~(\ref{eq:expansion}).
The question is whether geometric scaling is still valid 
in this case or
if there are scaling violations which could also be universal.
Another intriguing question is the meaning of the strong
absorption bound in the initial condition.
We think that the mathematical apparatus for nonlinear partial
derivative equations could provide interesting hints for these
questions.

As for the physical prospects of our approach, it is inspiring to
note that the same equation appears in many different contexts, among
which the problem of polymers on disordered trees
\cite{Derrida:88}.
In that context, $\beta$ plays the r\^ole of a temperature and 
there is a phase transition at $\beta\!=\!\beta_c$. $\beta\!>\!\beta_c$
corresponds to a phase of spin glass type.
Similar features have been discussed in multiparticle collisions
\cite{Bialas:1988wd}.
\\

\begin{acknowledgments}
R.P. wishes to thank B.~Derrida for introducing the subject of
traveling waves to him, 
and E.~Iancu for interesting discussions on saturation.
\end{acknowledgments}

%\bibliography{kov}

\begin{thebibliography}{10}
\bibitem{endnote18}{To see that the results coincide, one notes 
that 
$\mathaccent "7016 \gamma $ is the solution of the equation 
$\chi ^{\prime }(\gamma )/\chi (\gamma )=1/(\gamma -1)$ when $\chi (\gamma )$ 
is approximated by $\bar\chi(\gamma)$, {\it cf.}
Eq.(\ref {eq:expansion}).}
\end{thebibliography}
%\bibliographystyle{h-physrev3}

\end{document}